\documentclass[11pt]{article}
\usepackage{appb}
\def\la{\langle} 
\def\ra{\rangle} 
\def\be{\begin{eqnarray}} 
\def\ee{\end{eqnarray}}
\begin{document}
\title{CHIRAL DISORDER AND DIFFUSION \\OF LIGHT 
     QUARKS IN THE QCD VACUUM\thanks{Invited talk by MAN at the Workshop
on the Structure of Mesons, Baryons and Nuclei, Cracow, May 26-30, 1998.}
}
\author{Romuald A. Janik$^1$,
Maciej A.  Nowak$^{1,2}$,
G\'{a}bor Papp$^{3}$ and
Ismail Zahed$^{4}$%
\address{%
$^1$ Department of Physics, Jagellonian University, 30-059 Krakow, Poland.\\
$^2$GSI, Planckstr. 1, D-64291 Darmstadt, Germany\\
$^3$ Institute for Theoretical Physics, E\"{o}tv\"{o}s University,
     Budapest, Hungary\\
$^4$Department of Physics and Astronomy, Stony Brook, New York 11794, USA.%
}}

\maketitle

\begin{abstract}
We give a pedagogical introduction to the concept that light
quarks diffuse in the QCD vacuum following the spontaneous
breaking of chiral symmetry. By analogy with disordered 
electrons in metals, we show that the diffusion constant for
light quarks in QCD is $D=2F_{\pi}^2/|\la\bar{q}q\ra|$ which
is about $0.22$ fm. We comment on the correspondence between
the diffusive phase and the chiral phase as described by 
chiral perturbation theory, as well as the cross-over to the
ergodic phase as described by random matrix theory. The 
cross-over is identified with the Thouless energy 
$E_c=D/\sqrt{V_4}$ which is the inverse diffusion time in
an Euclidean four-volume $V_4$.
\end{abstract}
\PACS{64.60Cn, 11.30Rd, 12.38.Aw}
%{PACS numbers : 64.60Cn, 11.30Rd, 12.38.Aw}
\vspace*{2mm}

{\bf 1.}
In this talk and borrowing from our recent work~\cite{USDIS},
we explain using theoretical arguments how the QCD vacuum can
be viewed as a chirally disordered  medium, with chiral quarks 
diffusing through Euclidean four-space $V_4$ with diffusion 
constant $D=2 F_{\pi}^2/|\la\bar{q}q\ra|$.
 
In section~2  we define the basic concepts of diffusion, a 
phenomenon well known in 1,2,3- dimensional electronic systems, 
where the coherent description of electron waves fails at transport 
distances greater than the mean free path. In section~3  we  argue that 
a similar behavior in 4-dimensional Euclidean space  allows us
to interpret the result of multiple scatterings of originally coherent 
quark waves, in the form of a diffusion pole.
In section~4 we show how the diffusive  picture merges with the 
universal predictions of random matrix models, provided that
the eigenvalues of the Dirac operator are smaller than the
Thouless energy $E= D/\sqrt{V_4}$. In section~5 we comment on 
the generic character of this description in comparison
with chiral perturbation theory, the instanton model and 
lattice QCD simulations.  In section~6 we list  some plausible
extensions of our description to important issues in QCD.

{\bf 2.}
The typical distance of a diffusing particle grows as a function of
time as $r^2(t)=D t$, where $D$ is a diffusion constant.
One could therefore identify a time scale in which a particle is
diffusing through the sample of a linear size $L$ as $\tau_c=L^2/D$. 
The corresponding energy scale known as the Thouless energy, is therefore
$E_c=1/{\tau_c}=D/L^2$. For energies {\it lower} than the Thouless energy (times
greater than $\tau_c$), particle probes the whole volume of the system, 
defining in this way an ergodic regime. Decreasing the energy
yields a minimal scale of the order
of the mean quantum spacing $\Delta$, where the quantum 
description of the system becomes dominant. 
On the other side, for energies {\em larger} than the Thouless energy, a 
particle has time to probe only small domains of the whole volume. 
This is a diffusive regime. 
For even larger energies the diffusive regime breaks down, and we enter a
ballistic regime - simply times are so short, that the particle cannot
even travel the mean free path (m.f.p.) between elastic collisions.
This scale hierarchy is shown beneath.
%, and Fig.~1 shows 
%intuitive picture of collisions in various regimes.
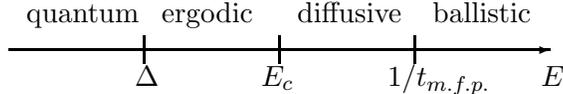
\begin{figure}[h]
\begin{center}
\setlength{\unitlength}{1.2mm}
\begin{picture}(60,10)
%\put(0,4){\line(1,0){60}}
\put(0,4){\vector(1,0){60}}
\put(15,2.5){\line(0,1){3}}
\put(30,2.5){\line(0,1){3}}
\put(45,2.5){\line(0,1){3}}
\put(14,0){\mbox{$\Delta$}}
\put(28,0){\mbox{$E_c$}}
\put(42,0){\mbox{$1/t_{m.f.p.}$}}
\put(59,0){\mbox{$E$}}
\put(2,7){quantum}
\put(17,7){ergodic}
\put(32,7){diffusive}
\put(47,7){ballistic}
\end{picture}
\end{center}
\end{figure}

Quantitatively,  the diffusion is easily described in terms of the
retarded and advanced Green's function $G^{\pm}(r,r',E)=\la r|(E-H \pm i\epsilon)^{-1}|r'\ra$, 
where $r,r'$ are positions in  d-dimensional space, and where the Hamiltonian  
$H$ contains the details of the disorder ({\it e.g.} $H$ could be 
an Anderson Hamiltonian). With the help of the Green's function,  
one could easily define \cite{EFETOV} the return probability $P(t)$ 
to the origin  in  fixed time $t$ for the diffusing particle with 
energy $E_F$. Specifically,
\be
P(t)=\frac{V}{2\pi \rho}\int d\lambda e^{-i\lambda t}
\la G^+(r,r,E_F+\lambda/2)G^-(r,r,E_F-\lambda/2)\ra
\label{ex1}
\ee
where $\rho$ is an average spectral density, and $\la...\ra$ is
the connected average over the underlying disorder. Ergodicity
implies that the disorder average and spectral average are the
same.

The process of averaging is equivalent to integrating over ``fast''
variables of the system. As a result (for details see   e.g.\cite{EFETOV})
the return probability is given by a classical formulae (in $t$ space and 
Fourier space, respectively)
\be
P(t)= \sum_Q e^{-DQ^2 t}   \quad ,\quad
P(\lambda)=\sum_Q \frac{1}{-i\lambda +D Q^2}
\label{ex2}
\ee
where the details of integration over the fast variables are hidden in the
diffusion constant $D$, and the diffusion modes $Q$ are ``slow''
(effective) degrees of freedom.  Since in the infinite volume limit
the solution (\ref{ex2}) reduces to the classical solution of the 
Helmholtz equation, the  effective ``slow'' variable is sometimes called 
a ``diffuson''. 

The diffusion pole could easily develop a gap, 
${-i\lambda\!+\!\!DQ^2}\!\rightarrow {-i\lambda\!+\!\!\gamma\!+\!DQ^2}$
if an exponential damping $\exp (-\gamma t)$ tantamount to loss of
coherence, multiplies $P(t)$. The damping is related to a finite coherence 
length $L_{coh}$ as 
\be
\gamma=D/L^2_{coh} \,.
\label{coh1}
\ee

In the ergodic regime (and ignoring the damping $\gamma$), the return 
probability is always 1. Technically, one could recover this result 
from (\ref{ex2}) by restricting the sum to {\em zero modes } Q=0.
In this way the results are space independent (infinitely long range, 
or equivalently, no derivatives), and therefore  belong to the same
universality
 class as 
ensembles of random matrices (i.e. a field theory in 0 dimensions). 
What only matters are the underlying
symmetries
of the Hamiltonian $H$, leading  in this way to the famous Dyson's threefold
path:
 Gaussian Unitary (GUE), Orthogonal (GOE) and Symplectic (GSpE) Ensembles,   
for hermitian, real and skew-symmetric Hamiltonians, respectively~\cite{MEHTA}.

{\bf 3.}
The idea that light quarks in the QCD vacuum might undergo (one-dimensional)
random walk was pointed out by Banks and Casher~\cite{BANKS}. It was later on
suggested in the context of the instanton liquid model~\cite{DIA,SHURYAK}  
that light quarks may behave like in a `semiconductor' 
depending on the instanton density in the vacuum. This point was established
using numerical simulations with instantons~\cite{SHURYAK,NVZ}. In this 
context it is suggestive to compare the Einstein-Kubo-Greenwood
formula for the conductivity $\sigma$ of a dc-current to the to the
Banks-Casher relation for the quark condensate as suggested
explicitly in~\cite{BOOK}
\be
\sigma= D\rho(E_F)  \,\,\,\,\,\,\,\,\,\,\,\,\leftrightarrow
\,\,\,\,\,\,\,\,\,\,\,\,\, |\la\bar{q}q\ra|=
\frac{1}{\pi V_4} \rho(0)
 \label{banks}
\ee
where $\rho(E_F)$ is the average electronic density at the Fermi level,
and $\rho(0)$ is the average spectral quark density of the eigenvalues
of the Dirac operator in QCD, averaged over all gluonic configurations.

What is new in our present arguments is that the analogy is indeed
quantitativaly realized in the QCD vacuum. Despite the fact that they
confine, light quarks diffuse in d=4 in the QCD vacuum~\cite{USDIS}.
The conductivity is played by the pion decay constant $F^2$, and the
Einstein-Kubo-Greenwood formula is precisely realized in QCD in
 the form of the (GOR)
Gell-Mann-Oakes-Renner relation~\cite{GOR}~\footnote{Recently, Stern~\cite{STERN} 
has presented arguments in which $F^2$ is also interpreted as a conductivity
but challenged the conventional form of the GOR relation and the 
Einstein-Kubo-Greenwood formula for light quarks.}. 
The idea that light quarks diffuse in
d=4 is key to understanding a number of phenomena in QCD in light
of results in disordered electronic systems. More importantly, it allows us
to organize certain aspects of infrared QCD as corrections to the Ohmic
conductivity, thereby providing a new and nonperturbative calculational scheme.
 
Indeed, the eigenvalue equation of the massless Dirac operator for fundamental
quarks in a fixed gluon field $A$ in Euclidean volume $V_4$ 
\be
i\nabla \!\!\!\!/ [A] \,q_k =\lambda_k [A] \, q_k
\label{01}
\ee
allows us to extend the theory into 4+1 dimension with proper time  $t$,
and define the probability $P(t)$ for a light quark to start at $x(0)$
in $V$ and return back to the same position $x(t)$ after a proper time
duration $t$. 
Here the four-dimensional  ``Hamiltonian'' $i\nabla \!\!\!\!/ [A]$ acts as 
a generator for the evolution operator along the additional time $t$, so 
the diffusion picture dwells in 4+1 dimensions.
 
We restrict our description to zero virtuality,
% {\it i.e.} $\lambda$ near  zero
{\it i.e.} we  focus on the diffusion of quarks corresponding to $E_F \sim
0$, by analogy with (\ref{banks}).
Therefore $P(t)$ reads~\cite{USDIS}, in analogy to (\ref{ex1})
\be
P(t) = \frac{V_4}{2\pi \rho}  \lim_{y\to x}{}
\int d\lambda  e^{-i\lambda t} 
\Big\la {\rm Tr}\left( S(x,y;z) S^{\dagger} (x,y; \bar{z})\right)\Big\ra_A
\label{des2}
\ee
with {\it complex} $z=m-i\lambda/2$, and Green's function (quark propagator)
\be
S(x,y; z) = \la x| \frac 1{i\nabla \!\!\!\!/[A] + iz} |y\ra \,.
\label{des3}
\ee
Here $\la...\ra$ denotes {\it connected} average over all gluonic
configurations in the QCD vacuum, making at first sight the r.h.s. 
untractable. Nevertheless,  it is possible  to relate the classical 
return probability to the properties of the pion, since
$S^{\dagger}(x,y;\bar{z})=-\gamma_5 S(y,x;z)\gamma_5$.
Therefore the return probability reads
\be
P(t) \sim  \lim_{y\to x}{}
\int  d\lambda 
\,e^{-i\lambda t} {\bf C}_{\pi} (x,y; z)
\label{des55}
\ee
where
\be
\,{\bf 1}^{ab}\, {\bf C}_{\pi} (x,y;z) = 
\Big\la {\rm Tr}\left(
S(x, y;z) i\gamma_5\tau^a S(y,x;z) i\gamma_5\tau^b\right)\Big\ra_A \,.
\label{des5}
\ee
We recognize that  ${\bf C}_{\pi}$ is the  analytically continued (with
$z=m-i\lambda/2$)
pion correlation function\footnote{The insertion of the isospin
generators $\tau^a$ is a formal `trick' to project onto  the 
connected part of the correlation function.}, given (for  
 $z=m$), due to  pion-pole dominance, by
\be
{\bf C}_{\pi} (x,y;m) \approx \frac{1}{V_4} \sum_Q e^{iQ\cdot (x-y)} 
\frac {\Sigma^2}{F_{\pi}^2} \frac 1{Q^2+m_{\pi}^2}
\label{des06}
\ee
with $Q_\mu =n_{\mu}2\pi/L$ in $V_4=L^4$ and $\Sigma=|\la \overline{q} q\ra |$.

Using the GOR~\cite{GOR}  relation
 $F_{\pi}^2m_{\pi}^2=m\Sigma$, 
and the analytical continuation $m\rightarrow z=m-i\lambda/2$, we 
find~\cite{USDIS}
\be
{\bf C}_{\pi} (x,y;z) \approx \frac{1}{V_4} \sum_Q e^{iQ\cdot (x-y)} 
\frac {2\Sigma}{-i\lambda + 2m + DQ^2}
\label{des6}
\ee
with the diffusion constant $D=2F_{\pi}^2/\Sigma$, in full analogy to (\ref{ex2}).
Inserting (\ref{des6}) into (\ref{des55}), and using the Banks-Casher
relation we conclude, after a contour integration that indeed
\be
P(t) = e^{-2mt}\sum_Q e^{-DQ^2 t} \,.
\label{des7}
\ee

For very large current quark masses $m$, the diffusion is suppressed.
For small $m$, we see that twice the current quark mass plays the role of the
damping $\gamma$,  introducing a cutoff on infinitely long diffusion paths.
For $m\neq 0$, the coherence length corresponding to the collective
excitations of the QCD vacuum is finite,  $L_{coh}=1/m_{\pi}$, 
and we see that this is again consistent with the GOR formula 
({\it cf.} (\ref{coh1})  with $D=2F_{\pi}^2/\Sigma$). 
The analytically continued pion propagator plays the role of the 
``diffuson'' degrees of freedom, after integrating the QCD fast variables.
The effective lagrangian for pions is  organized on the basis 
of chiral counting~\cite{GL}, and similarly effective lagrangians 
for diffusons are basically sigma models~\cite{EFETOV}. 

It is now easy to identify all the relevant length scales defined in section~2
and separating the different regimes of the QCD vacuum viewed as a disordered
medium. The scale which separates the quantum regime from the ergodic one 
is the average level spacing $\Delta$ from the spectral function of the
Dirac operator, $\Delta=\pi/\Sigma V_4$. The Thouless energy equals 
$D/\sqrt{V_4}$  with $D=2F^2_{\pi}/\Sigma$. In this way we define the ergodic 
regime. The diffusive regime starts to be relevant for eigenvalues of the Dirac
operator greater than the Thouless energy. The diffusive regime merges with 
the ballistic regime when the eigenvalues approach twice the value of the
{\it constituent} quark mass. Indeed, if the propagation in time is not enough
to cover a mean free path, the concept of ``dressing'' the quark through 
multiple  collisions  becomes obsolete. We note that the segregation of 
scales take place naturally in a {\it finite} Euclidean volume $V_4$, hence
the usefulness of the `box'~\cite{USDIS}.

{\bf 4.}
It is inspiring to see how the universal (ergodic) regime 
appears as a limit of the diffusive regime. This regime 
gained recently a lot of attention in `random' QCD, due to several 
remarkable agreements between predictions based on chiral random matrix models 
and lattice QCD simulations. Like in disordered metallic systems, where 
the ergodic regime appears as a consequence of restricting to zero 
modes of diffusion, a similar simplification  operates in QCD.
For energy scales {\it smaller} than the Thouless energy, the
quark return probability is equal to 1, which again corresponds to 
the zero-mode approximation. Since restricting to $Q=0$  is equivalent 
to keeping only constant modes (no derivatives), much of the QCD dynamics 
becomes irrelevant.
The partition function for {\em full} QCD in this regime is simply~\cite{GL}
\be
Z(m)=\int dU e^{\Sigma V_4 {\rm Tr}\, (m (U+U^{\dagger}))} \,.
\label{gl}
\ee 
One sees that the r.h.s. depends {\it solely} on the way  chiral
symmetry is spontaneously broken (the integration $dU$ over the 
coset space of Goldstone modes)  and on the explicit pattern  of 
breaking chiral symmetry (exponent), here in the 
$(N_f,\bar{N_f}) +(\bar{N_f},N_f)$ representation.
Like in the nonchiral case, here also only three generic 
scenarios are possible,
depending on the symmetries of the Dirac operator~\cite{VER}. 
For QCD with $N_c \ge 3$ where quarks are in fundamental representation,
theory belongs to the universality class of Chiral GUE.
For QCD with $N_c=2$ the theory belongs to the universality class of 
Chiral GOE. For QCD with quarks in the adjoint representation the theory
belongs to the universality class of Chiral GSpE.

The ``chirality'' (off-block diagonal structure) of the random ensembles
is the remnant of the chiral property $[iD\!\!\!\!/, \gamma_5]_+=0$.
All three possibilities have a simple interpretation from the point of
view of approaching the ergodic regime from the diffusive regime~\cite{USDIS}. 
For QCD with three or more colors, each quark orbit is traversed once.
For QCD with two colors, each quark orbit is traversed twice,
since  $SU(2)$ does not `distinguish' between quarks and antiquarks.
Finally, the adjoint representation of quarks means that we  have  
supersymmetric QCD (same group representation for quarks and gluons). 
Intuitively,  in  this case the quark traverses only ``half'' of the
orbit, due to the fact that supersymmetry is ``taking a square root''
from the Dirac equation. More formally, in the diffusive regime of QCD 
one could postulate following semiclassical arguments 
from classical chaos~\cite{CLASSCHA}
\be
K(t)\approx 2t\Delta^2/(4\pi^2 \beta) P(t)
\label{Guz}
\ee
where the spectral form factor 
$K(t)$ is directly related to two-level quantum correlation function, 
and $\beta=1,2,4$ for Chiral GUE,GOE,GSpE, respectively.

To summarise: we would like to stress once more, that several nontrivial 
results for QCD below the Thouless energy, like the existence of 
the Leutwyler-Smilga
sum rules~\cite{LS}, the existence of universal microscopic correlators
suggested by the Stony Brook group~\cite{SB} and proven to be universal by the 
Copenhagen group~\cite{POUL}, the observation of universal  oscillations in lattice
spectra~\cite{WETTIG} and other results,  are all various facets of the same 
fact that ``$P(t)=1$ in the ergodic regime.''

{\bf 5.}
The diffusive picture of the QCD vacuum, sketched here on the basis 
of  analogies to condensed matter physics, has several relations to 
existing descriptions of the spontaneous breakdown of chiral
symmetry. In the language of chiral power counting, the ergodic regime 
corresponds to the limit $mV_4 \sim 1$, whereas the diffusive regime to the 
limit $m^2 V_4 \sim 1$. The first counting dwarfs  the contribution of the 
non-zero modes to the pion propagator ${\bf C}_{\pi}$, leaving only the
zero mode. The second counting enhances the contribution of the non-zero modes
over the zero mode, leading to standard chiral perturbation theory~\cite{GL}.

It is interesting to compare the diffusive picture to the model of
instantons, since the analytical scenario put forward by Diakonov and 
Petrov~\cite{DIA} and numerical one suggested by Shuryak~\cite{SHURYAK} are 
(in our  knowledge) the first detailed realization of some of the ideas on 
disorder in a QCD model. In the instanton model, the average
 $\la...\ra$ over all gluonic configurations is performed explicitly,
 assuming an ansatz for a random model of instantons and antiinstantons.
Disorder comes from random positions and colors of instantons, and
chirality is built in through the chiral properties of 
(right/left) fermionic zero modes for each instanton (antiinstanton).
Since in the instanton model one could express parametrically each 
dimensionfull quantity in terms of the average instanton radius $\bar{r}$
and average inter-instanton distance $R\sim 1$ fm,
the smallness of the diffusion constant   $D=0.22$ fm reflects simply 
the diluteness of the instanton vacuum, where $\bar{r}/R=0.2-0.3$. The fact 
that the instanton vacuum has a regime consistent
 with  (\ref{gl}) is well known.  
A recent numerical study by Osborn and 
Verbaarschot~\cite{OV} has also confirmed the existence of the Thouless energy 
in this model, as we originally predicted~\cite{USDIS}. 

Perhaps the most interesting comparison of our predictions is a direct 
lattice simulation. From (\ref{Guz}), a direct comparison to the lattice 
is possible since the l.h.s. is explicitly measured on the lattice, and the
r.h.s. is explicitly calculable for all regimes of disorder.
The first interesting investigation confirming some of the results presented 
here  was recently carried out by Berbenni-Bitsch {\it et al.}~\cite{WETTIG2} 
In particular, they have considered the dimensionless ratio $\lambda_{Max}/\Delta$, 
where $\lambda_{Max}$ is the {\em maximal} eigenvalue of the Dirac operator 
for which the universal predictions based on random matrix theory still hold. 
Since $\lambda_{Max}$ is nothing but the Thouless energy, this dimensionless
ratio reads
\be
\frac{E_c}{\Delta}=\frac{2}{\pi}F_{\pi}^2 L^2
\label{weid}
\ee
where we used the Banks-Casher relation and the preceding definitions 
for the Thouless energy $E_c$ and the diffusion constant $D$, respectively.
By studying various lattice sizes, the authors~\cite{WETTIG2} 
have observed that the scaling is consistent with (\ref{weid}) and 
identified the Thouless energy for the lattice.

{\bf 6.}
The present ideas on chiral disorder translate  concepts
of disorder from  condensed matter physics to low-energy chiral QCD. 
Most of the results are directly amenable to 
lattice studies.\footnote{We also hope that the ballistic regime 
is amenable  to a lattice investigation despite the proximity of this
regime to the UV-spectrum.} A number of non-perturbative investigations can 
be carried out in the present context. In particular, \\
- the effects of finite temperature and chemical potential 
  on the diffusive properties of the QCD vacuum;\\
- the role of the number of flavors and finite $\theta$ (strong CP violating)
  angle on the diffusion scenario;\\
- the change of the spectral properties of the `diffuson' at the 
  critical temperature and the dependence of the return probability on 
  the critical exponents in QCD phase transitions;\\
- the modifications of the diffusion due to the addition of
  electromagnetic   and chromomagnetic external fields 
  (negative magnetoresistance of the QCD
   vacuum, Bohm-Aharonov like effects due to the presence of fluxons
   or maximally abelian-projected monopoles).\\
Some of these issues will be brought up next.

\vskip 0.5cm
MAN thanks the organizers of the Workshop on the Structure of Mesons,
Baryons and Nuclei  and the conference Meson '98 for an invitation.
This work was supported in part by the US DOE grant DE-FG-88ER40388, by the 
Polish Government Project (KBN) grants 2P03B04412 and 2P03B00814 and by the 
Hungarian grants FKFP-0126/1997 and OTKA-F026622.

\end{document}